\documentclass[twocolumn,showpacs,aps,prb,amsmath,amssymb,floatfix]{revtex4}
\usepackage{graphicx}
\usepackage{dcolumn}
\usepackage{bm}
\begin{document}
\title{Disorder-driven superconductor-normal metal phase transition
in quasi-one-dimensional organic conductors}
\author{E. Nakhmedov$^{1,2}$ and R. Oppermann$^{1,3}$}
\affiliation{ $^1$Institut f\"ur Theoretische Physik,
Universit\"at W\"urzburg,
D-97074 W\"urzburg, Germany\\
$^2$Institute of Physics, Azerbaijan National Academy of
Sciences, H. Cavid str. 33, AZ1143 Baku, Azerbaijan\\
$^3$Institut de Physique Th\'eorique, CEA Saclay, Orme des Merisiers, F-91191 Gif-sur-Yvette, France}
\date{\today}
\begin{abstract}
Effects of non-magnetic disorder on the critical temperature
$T_c$ and on diamagnetism of quasi-one-dimensional superconductors
are reported.
The energy of Josephson-coupling between wires is
considered to be random, which is typical for dirty organic
superconductors. We show that this randomness destroys phase
coherence between wires and that $T_c$ vanishes discontinuously at
a critical disorder-strength. The parallel and transverse components
of the penetration-depth are evaluated. They
diverge at different critical temperatures $T_c^{(1)}$ and $T_c$,
which correspond to pair-breaking and phase-coherence breaking
respectively.
The interplay between disorder and quantum phase fluctuations is
shown to result in quantum critical behavior at $T=0$, which
manifests itself as a superconducting-normal metal phase transition
of first-order at a critical disorder strength.
\end{abstract}
\pacs{74.78.-w, 74.62.-c, 74.70.Kn, 74.50.+r}
\maketitle
\section{Introduction}
Although more than a quarter of a century has elapsed since the
discovery of superconductivity in the quasi-one-dimensional
(quasi-1D) organic charge-transfer (Bechgaard) salts of
$(TMTSF)_2X$-type (where $TMTSF$ stands for
tetramethyltetraselenofulvalinium and $X = PF_6, ClO_4,NO_3$ being
a strong electron acceptor or anion) \cite{jmrb80,bj09}, many
distinct properties of this material still remain a matter of
debate. Among those one may mention the pairing symmetry, the
remarkable sensitivity of the critical temperature to irradiation
\cite{cchh82,bouf82}, large upper critical field $H_{c2}$ etcetera
(see, also Ref.\onlinecite{powell08}). The irradiation destroys
anion order, introducing thus non-magnetic damage into system that
led Abrikosov to suggest the possibility of triplet pairing
\cite{aaa83} in the organic salts. Nevertheless, the experimental
evidence \cite{skzp07} that the spin-susceptibility decreases at
low magnetic fields in the superconducting (SC) state of the
$(TMTSF)_2ClO_4$ organic conductor, disfavors the triplet pairing
mechanism and supports spin-singlet pairing.

Low temperature properties of organic superconductors are known
to be very sensitive to disorder. Alloying with anions,
$x$-ray irradiation, or cooling rate controlled anion reorientation
introduces non-magnetic randomness into the system, however leaving
unchanged, to a large extent, the backbone structure and the unit cell
of the organic superconductors. There is a common agreement that disorder, when introduced
by means of these experimental methods, must be characterized as
non-magnetic. Yet, it was shown \cite{cchh82, iys98,japm04,japj05} to suppress the SC phase. It is worth noting
that the methods of generating non-magnetic disorder in layered
organic superconductors are similar (see, e.g. Ref.\onlinecite{pk04})
to those in quasi-1D systems, and, therefore a similar mechanism of
SC state suppression in these two classes of materials may exist.

Studies of disorder effects on the superconducting phase have a
long history. The superconducting transition temperature $T_c$ for
$s$-wave pairing has been shown to be insensitive to the
scattering rate on non-magnetic impurities, which is known as the
Anderson theorem \cite{anderson59}. In contrast to non-magnetic
impurities, paramagnetic impurities break time-reversal symmetry
of the $s$-pairing, and suppress the SC-phase \cite{ag61} at some
critical concentration of the impurities. Correlation between the
paramagnetic impurities via the Ruderman-Kittel-Kasuya-Yosida
(RKKY) interaction yields a spin-glass phase below $T=T_g$, which
was shown \cite{gl02} to shift the superconducting transition
point towards higher temperatures. The Anderson theorem is not
applicable to unconventional superconductors with $d$-wave pairing
symmetry \cite{tk00}. A small concentration of non-magnetic
impurities may destroy $d$-wave pairing, producing a finite
lifetime for quasi-particles near the nodes in the gap
\cite{lee93,hg93,ss89}.

Inter-chain (inter-layer) pairings as well as intra-chain
(intra-layer) pairings, occurring at different local critical
temperatures in quasi-1D (quasi-2D) systems, yield also an
inhomogeneous nodal order parameter \cite{nt94,nh98}, which
affect considerably the upper critical magnetic field.

Suppression of superconductivity in the presence of non-magnetic
impurities can in general be realized by destroying either
the modulus or the phase coherence of the order parameter.
Interplay between superconductivity and Anderson localization in a
strongly disordered superconductor was shown \cite{ml85, ro90,
grt01, szh01, gl01, swt04, sok08, mrms08, mpss08,zp86} to result in
spatial inhomogeneity of the order parameter. Diffusive scattering
of particles in the random field of non-magnetic impurities
enhances Coulomb repulsion \cite{fink87,amr83}, and
consequently, reduces the amplitude of the order parameter.
Mesoscopic fluctuations in a superconducting thin film were also
shown \cite{sf05} to yield a spatial inhomogeneity of the order
parameter.

Effects of order parameter phase fluctuations on the
superconducting transition temperature $T_c$ have been
studied in Refs.\onlinecite{el74,fy76,sb83,ek95,nf98}. It is well known that
there is no SC phase transition in one-dimensional (1D) and
two-dimensional (2D) systems \cite{rice65}, since strong
fluctuations of the order parameter phase destroy off-diagonal
long-range order (ODLRO) in a single SC wire and in an isolated SC
film. Even a small interchain-coupling in clean quasi-1D
superconductors restores however ODLRO together with a finite
transition temperature. The suppression of $T_c$ by strong phase
fluctuations in clean quasi-1D superconductors was analyzed in
Ref.\onlinecite{el74,fy76}. Classifying superconductors with small stiffness as bad
metals, Emery and Kivelson have evaluated \cite{ek95} a critical
temperature of phase ordering by formally dividing a clean bulk
superconductor into small regions with well defined phase, and
have shown considerable suppression of an SC-phase by phase fluctuations.
Effects of disorder on phase fluctuations were however not
considered in all of these papers.

Distinct structural peculiarities of quasi-1D organic
superconductors demand a special theory, which should take into
account non-magnetic randomness as well as phase fluctuations.
Indeed, the high purity of the organic superconductor backbone,
even in an overall dirty limit, excludes spatial inhomogeneity
of the order parameter modulus along the SC-wires. This renders
inapplicable the above-mentioned theories for pair breaking.

In contrast to these previous activities we study in this article
a disorder-driven superconductor-normal metal phase transitions
due to the competition of non-magnetic randomness and
phase-fluctuations in quasi-1D superconductors. We consider weakly
linked quasi-one-dimensional superconductors with random
Josephson-couplings between pure one-dimensional (1D) SC wires.
Singlet pairing is assumed within the wires.
Therefore, we assume that non-magnetic randomness does not affect
the order parameter amplitude.

We demonstrate in this article that

(i) non-magnetic randomness in the Josephson-coupling destroys
correlation of the phases between different chains in quasi-1D
superconductors even in the classical phase fluctuation regime,

(ii) non-magnetic randomness yields quantum critical behavior in
addition. A superconducting normal-metal phase transition occurs
at $T=0$ increasing the strength of disorder, and

(iii) a suppression of the SC phase occurs discontinuously as well
both the classical and the quantum phase fluctuation regimes as a
first-order phase transition when the disorder-strength reaches
a critical value. We derive parallel and perpendicular components
of the penetration depth, $\lambda_{\|}$ and $\lambda_{\perp}$,
and show that they diverge at different critical temperatures
$T_c^{(1)}$ and $T_c$, which correspond to pair-breaking in
the wires and to phase coherence breaking between the SC wires,
respectively.

The paper is organized as follows: In Sec. II we study the
interplay of randomness in the Josephson energy with phase
fluctuations inside the classical regime. Although an arbitrary
small interchain coupling in a clean quasi-1D superconductor
stabilizes the ODLRO giving a finite transition temperature $T_c$,
we show in this Section that non-magnetic disorder in the
Josephson energy suppresses $T_c$ discontinuously when the
disorder-strength reaches a critical value. In Sec. III, we
investigate effects of randomness on the transverse rigidity and
on $T_c$ in the quantum fluctuations regime. We show in this
Section that a suppression of the SC phase is managed by two
parameters characterizing the disorder-strength and a dynamical
charging parameter $\alpha$ in the system. The quantum criticality
at $T = 0$ and the phase transition at finite temperature are
described separately. The transverse rigidity in the field of the
phases is shown to vanish discontinuously. The jump at the
breakdown point decreases monotonically with increasing $\alpha$
in the interval of $0 \le \alpha < 1$, and vanishes for $\alpha =
1$. The breakdown point is pushed towards higher values of the
disorder-strength as $\alpha$ increases. We show that, under a change
of the disorder-strength, the critical temperature evolves similarly
as the transverse rigidity at $T=0$. The analysis of diamagnetism
in this Section reveals completely different behavior for parallel
and perpendicular components of the penetration depth.
The penetration depth for a magnetic field parallel to the SC-wires
is shown to be defined by the phase-phase correlator between
two neighboring wires, which behaves non-linearly in temperature
and reveals a discontinuous dependence on disorder; by contrast
the perpendicular penetration depth does not depend on disorder
and shows a conventional temperature dependence. In Sec. V we
summarize results obtained in the paper. The explicit evaluation
of the phase-phase correlator is given in the Appendix.
%
%
\section{Classical phase fluctuation regime}
A quasi-1D superconductor is modeled as a system of one-dimensional wires,
which are placed regularly and parallel to each other, forming for example
a square lattice in the cross-section.
Weak tunneling between the chains results in an open Fermi surface
for the normal metallic state, and yields also the Josephson-coupling
between nearest-neighboring chains in the superconducting state and
strong anisotropy in kinetic properties.

The free energy functional of a quasi-1D superconductor, which is weakly
linked with Josephson-coupling energy $E_{\bf j, j+g}$ between
nearest-neighbor chains, can be written in the presence of the magnetic
field ${\bf B}$ as
\begin{eqnarray}
F_{st} &=& N_s^{(1)}(T) \xi_{\|} \sum_{\bf j} \int dz \bigg
\{\frac{\hbar ^2}{8 m_{\|} \xi_{\|}^2} \bigg( \frac{\partial
\varphi_{\bf j}}{\partial z} - \frac{2e \xi_{\|}}{\hbar c}
A_z\bigg)^2 +
\nonumber\\
&&\hspace{-1cm}\sum_{{\bf g}=\pm 1}E_{\bf j,j+g} \left[1 - \cos \big(
\varphi_{\bf j} - \varphi_{\bf j+g} + \frac{2e \xi_{\|}}{\hbar
c}\int_{\bf j}^{\bf j+g} {\bf A}_{\perp} d {\bf r}_{\perp} \big)\right]
{}\nonumber\\
&+& \xi_{\|} a^2_{\perp} \frac{({\bf B}({\bf r}) - {\bf
B}_{ext})^2}{8 \pi} \bigg \}, \label{freeenergy}
\end{eqnarray}
where $\varphi_{\bf j}(z)$ denotes the phase of the order
parameter $\Delta_{\bf j}(z) = |\Delta_{\bf j}| \exp (i
\varphi_{\bf j}(z))$ at a point with dimensionless coordinates
${\bf r}=\{{\bf j} = \{ j_x, j_y \}, z \}$, ${\bf A} = \{{\bf
A}_{\perp}, A_z \}$ is the vector-potential, and $N_s^{(1)}$ the
linear density of SC electrons with $N_s^{(1)}(T)= N_s^{(1)}(0)
\equiv N_N^{(1)} \simeq \frac{p_F}{\hbar}$ at $T \ge T_c^{(1)}$,
and $N_s^{(1)}(T) = N_s^{(1)}(0) \tau(T)$ with $\tau (T) =
\frac{T_c^{(1)} - T}{T_c^{(1)}}$ at $T \le T_c^{(1)}$.
Dimensionless coordinates ${\bf r} = \{ {\bf j}, z \}$ are
introduced on the scale of the longitudinal $\xi_{\|} = \frac{\hbar^2
N_s^{(1)}(0)}{4 m_{\|} T_c^{(1)}}$ and the transverse $\xi_{\perp}
\sim a_{\perp}$ components of the coherence length, so ${\bf r} =
\{ r_{\|} = \{j_x a_{\perp}, j_y a_{\perp} \}, z \} \to \{ {\bf j}
= \{j_x, j_y \}, z/ \xi_{\|} \}$. The last term in
Eq.(\ref{freeenergy}) describes the Josephson-coupling with the
coupling energy $E_{\bf j, j+g}$ between the wires, which is
minimal for a coherent tunneling $(\varphi_{\bf j}(z) \sim
\varphi_{\bf j+g}(z))$ of Cooper pairs between neighboring
wires.

Fluctuations of the order parameter modulus can be neglected for
pure superconductors \cite{el74} far from $T_c^{(1)}$, satisfying
the condition $(T_c^{(1)} - T )/T_c^{(1)} \gg n^{-2/3}$, where
$T_c^{(1)}$ is the mean-field critical temperature calculated for
an isolated wire and $n$ is the number of bands in each chain
\cite{Pieirls}. Therefore, the contributions to the free energy
functional (\ref{freeenergy}), coming from the modulus of the
order parameter $| \Delta_{\bf j}|$, are omitted.

We assume the Josephson energy $E_{\bf j, j+g}$ to be a random
parameter with Gaussian distribution, centered at the mean value
$E_{\bf g}$, given by
\begin{equation}
P \{ E_{\bf j, j+g} \} = \frac{1}{\sqrt{2 \pi W^2}} \exp \big \{ -
\frac{(E_{\bf j, j+g} - E_{\bf g})^2}{2 W^2} \big \}.
\label{gauss}
\end{equation}
The variance $W^2$ is taken as a measure of disorder strength
in this coupling of nearest neighbor chains.
Employing the replica trick one can calculate the average value of
the free energy $\mathcal{F} =-T \langle \ln Z \rangle$ over
disorder. As usual we use $\langle \ln Z \rangle = \lim_{n \to 0}
\frac{\partial}{\partial n}\langle Z^n \rangle$ and, in addition,
express the n-th power of the partition function $Z = \int
\prod_{\bf j} \mathcal{D}\varphi_{\bf j} e^{-F_{st}/T}$ by means
of replicated fields $\varphi^a$, $a=1...n$, as (for ${\bf B}=0$)
\begin{eqnarray}
\hspace{-0.0cm} \langle Z^n \rangle = \int \prod
\mathcal{D}\varphi_{\bf j}^a \exp \bigg \{-\frac{N_s \xi_{\|}}{T}
\sum_{{\bf j},a}\int dz \bigg[\frac{\hbar^2}{8m_{\|}\xi_{\|}^2}
\left(\frac{\partial
\varphi_{\bf j}^a}{\partial z}\right)^2 \hspace{-0.2cm} \nonumber\\
+\sum_{\bf g} E_{\bf g}[1-\cos(\varphi_{\bf j}^a-\varphi_{\bf
j+g}^a)]\bigg] + \frac{1}{2} \sum_{\bf j,g}\bigg[\frac{N_s
\xi_{\|}W}{T}\times \hspace{+1.0cm} \nonumber\\
\times \int dz\sum_{a} [1-\cos (\varphi_{\bf j}^a - \varphi_{\bf
j+g}^a)]\bigg]^2 \bigg \}, \hspace{+3.0cm}
\end{eqnarray}
The quadratic cosine term is linearized with the help of a
Hubbard-Stratonovich transformation by introducing an auxiliary
field $\zeta_{\bf j,g}$. As a result, the sum over the replica
variable $a$ is factorized and the replica limit can be performed,
yielding for the averaged free energy
\begin{eqnarray}
\mathcal{F} = -T \int \prod_{\bf j, g}
\frac{N_s^{(1)}\xi_{\|}}{\sqrt {2 \pi}}d \zeta_{\bf j, g} e^{-
\frac{N_s^{(1) 2}\xi_{\|}^2}{2}\zeta_{\bf j, g}^2} \times \nonumber\\
 \times \ln \int \prod \mathcal{D} \varphi_{\bf j}
e^{-F/T}\hspace{+3.2cm}\\
{\rm with}\quad F = N_s^{(1)}\xi_{\|} \sum_{\bf j} \int dz \bigg\{
\frac{\hbar^2}{8 m_{\|} \xi_{\|}^2} \left( \frac{\partial
\varphi_{\bf j}}{\partial z} \right)^2 +
\nonumber\\
\hspace{-2.6cm} \sum_{\bf g} (E_{\bf g}-
N_s^{(1)}\xi_{\|}W\zeta_{\bf j,g}) [1 - \cos(\varphi_{\bf j}
-\varphi_{\bf j+g})] \bigg\}. \label{eff-energy}
\end{eqnarray}
The average value of a given functional $C(\{\varphi_{\bf j} \})$,
e.g. $\cos \varphi_{\bf j}$ or $\cos(\varphi_{\bf j}-\varphi_{\bf
j+g})$, can be obtained according to the relation $\langle\langle
C(\{\varphi_{\bf j} \})\rangle\rangle =-T\frac{\delta}{\delta
\eta_{\bf j}}\langle \ln Z\rangle |_{\eta_{\bf j}=0}$ by adding
the source term $\sum_{\bf j}\int dz \eta_{\bf j} C(\{\varphi_{\bf
j} \})$ to the free energy functional, which yields for the
correlator
\begin{eqnarray}
&&\langle \langle C(\{ \varphi_{\bf j} \})\rangle \rangle = \int
\prod_{\bf j, g} \frac{N_s^{(1)}\xi_{\|}}{\sqrt {2 \pi}}d
\zeta_{\bf j, g} e^{- \frac{N_s^{(1) 2}\xi_{\|}^2}{2}\zeta_{\bf j,
g}^2} \times \nonumber\\
&&\times \frac{\int \mathcal{D} \varphi C (\{\varphi_{\bf j}\})
e^{- F/T}}{\int \mathcal{D} \varphi e^{- F/T}}, \label{av}
\end{eqnarray}
where the double bracket $\langle \langle \dots \rangle \rangle$
means averaging over thermodynamic fluctuations and over
randomness. In order to estimate an asymptotic behavior of the
correlator, e.g. $\langle \langle \cos(\varphi_{\bf j} -
\varphi_{\bf j+g}) \rangle \rangle$ we write the integrand of
Eq.(\ref{av}) as $\exp \{-N_s^{(1) 2}\xi_{\|}^2 f(\zeta_{\bf
j,g})\}$, and apply the stationary-phase approximation to
determine an extremal value of the auxiliary field $\bar
{\zeta}_{\bf j,g}$ minimizing the function $f(\zeta_{\bf j,g})$.
The minimal value of $\zeta_{\bf j,g}$ is obtained to be
\begin{eqnarray}
\bar{ \zeta}_{\bf j, g} =\frac{W}{T}\int dz \bigg
\{\langle\cos(\varphi_{\bf j}(z)-\varphi_{\bf j+g}(z))\rangle -
\nonumber\\
- \frac{ \langle \cos(\varphi_{\bf j}(z)-\varphi_{\bf j+g}(z))
\cos(\varphi_{\bf j}(0)-\varphi_{\bf j+g}(0))\rangle}{\langle
\cos(\varphi_{\bf j}(0)-\varphi_{\bf j+g}(0))\rangle} \bigg \}.
\end{eqnarray}
The constant $N_s^{(1)}\xi_{\|}$ on the exponent can be estimated
to be equal to $N_s^{(1)}\xi_{\|}\simeq
\frac{\epsilon_F}{T_c^{(1)}} \sim 10^3$ for the organic
superconductors with $\epsilon_F$ being the Fermi energy, which
ensures a sharply peaked saddle point of the integrand. The
thermodynamic averages in the expression of $\bar{\zeta}_{\bf
j,g}$ are taken with the free energy functional, given by
Eq.(\ref{eff-energy}), at the saddle point $\zeta_{\bf j,
g}=\bar{\zeta}_{\bf j, g}$. So, a contribution of the non-magnetic
randomness to the effective free-energy functional is proportional
to the variance of the phase correlator, which gives an idea on
the form of the disorder-dependent term in the effective
functional. Note also that the saddle point for the averaged free
energy $\mathcal{F}$ is given as $\tilde{\zeta}_{\bf j,g} = -
\frac{W}{\mathcal{\bar F}} \int dz \langle [1-\cos(\varphi_{\bf
j}(z)-\varphi_{\bf j+g}(z))] \rangle$, where $\mathcal{\tilde F}$
is the value of the free energy at the saddle point.

The critical temperature for the quasi-1D superconductors can now
be found from Eq.(\ref{av}),written for $\cos\varphi_{\bf j}$ by
using the self-consistent mean-field method \cite{el74}, which
consists in replacing the phase correlations of the cosine term by
\begin{eqnarray}
&\sum_{\bf g}& E_{\bf g} [1-\cos(\varphi_{\bf j}(z)-\varphi_{\bf
j+g}(z))] \nonumber\\
&\longrightarrow&
E_{\perp} [1 - \langle \langle \cos(\varphi) \rangle \rangle \cos
(\varphi (z))],
\end{eqnarray}
where $E_{\perp}  = \sum_{\bf g} E_{\bf g}$. For a clean system
$\langle \langle \cos(\varphi) \rangle \rangle_{eff}$ was chosen
\cite{el74} to be equal to $\langle \cos(\varphi) \rangle$. For
the disordered superconductor we choose $\langle\langle \cos
\varphi \rangle\rangle_{eff}= \langle\langle \cos \varphi
\rangle\rangle - N_s^{(1)}\xi_{\|} \frac{\langle\langle \cos
\varphi \rangle^2\rangle-\langle\langle \cos \varphi
\rangle\rangle^2}{\langle\langle \cos \varphi \rangle\rangle}$.
The functional integral over the phases in Eq.(\ref{av}) can not
yet be evaluated, even after this simplification. Taking advantage
of the smallness of $(E_{\perp}- N_s^{(1)}\xi_{\|} W \zeta)
\langle\langle\cos(\varphi)\rangle\rangle_{eff}$ near $T_c$, we
expand both the numerator and the denominator of the integrand of
Eq.(\ref{av}), written for $\langle \langle \cos(\varphi) \rangle
\rangle_{eff}$, in this parameter. The thermodynamic averages
become pure one-dimensional after this expansion, which can be
taken easily, yielding a power series of $\zeta$ for the
integrand. Therefore, the integration over $\zeta$ is immediately
performed. Since all higher order in $\langle \langle
\cos(\varphi) \rangle \rangle_{eff}$ terms of the expansion vanish
at $T=T_c$, we get the equation for $T_{c}$
\begin{eqnarray}
\hspace{-2cm} 1 &=& \frac{E_{\perp} N_s^{(1)}
\xi_{\|}}{T_c}\biggl(1 - \frac{W^2
\xi_{\|} N_s^{(1)} \eta ^2}{ T_c E_{\perp}}\biggr)\times\nonumber\\
&&\hspace{1cm}\times \int \langle \cos(\varphi(0))
\cos(\varphi(z))\rangle dz, \label{Tc}
\end{eqnarray}
where $\eta$ is the coordination number. The phase-phase
correlator in Eq.(\ref{Tc}) is calculated in the clean limit of
the $1D$ free energy functional, obtained from
Eq.(\ref{freeenergy}) by setting $E_{\bf j,j+g} = 0$, which
returns (see for example Ref. \cite{rice65})
\begin{equation}
\langle \cos (\varphi (0)) \cos(\varphi (z)) \rangle = \exp
\{-|z|/r_c \}, \label{1Dcorrelator}
\end{equation}
where $r_c = \hbar^2 N_s^{(1)}(T)/2 m_{\|} \xi_{\|}T$.
Introducing a dimensionless $T_c$-shift by
\begin{equation}
t = \sqrt{\eta \epsilon_F E_{\perp}}
\left(\frac{1}{T_c} - \frac{1}{T_{c}^{(1)}} \right),
\end{equation}
with $\epsilon_F$ being the Fermi energy, and a dimensionless disorder parameter
\begin{equation}
q = \frac{W^2}{E_{\perp}} \sqrt{\frac{2 m_{\|}
\xi_{\|}^2 \eta}{\hbar^2 E_{\perp}}} = \frac{ W^2}{2 E_{\perp}
T_c^{(1)}} \sqrt{\frac{\eta \epsilon_F}{E_{\perp}}},
\label{q-eq}
\end{equation}
Eqs.(\ref{Tc}), (\ref{1Dcorrelator}) yield
\begin{equation}
1 = t^2 (1 -q t).
\label{Tc3}
\end{equation}

\begin{figure}
\resizebox{.48\textwidth}{!}{%
\includegraphics[width=1cm]{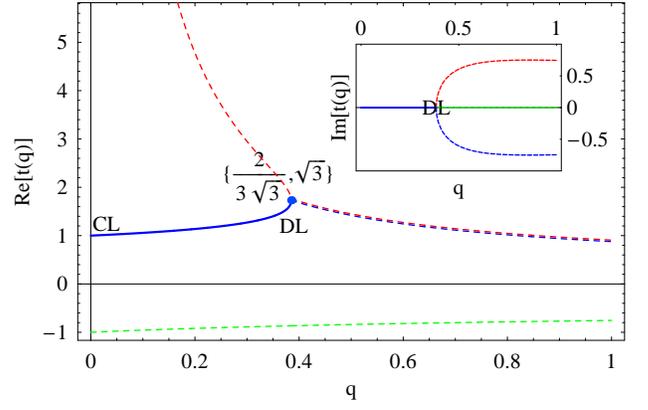}}
\caption{The physical solution $t(q)$, giving the
$T_c(W)$-dependence, within the full range from clean limit (CL:
$q=0$) to the dirty limit (DL: $q_c=2/3\sqrt{3})$) is highlighted
as the bold (blue) curve. Formal solutions of the cubic
Eq.(\ref{Tc3}) are shown for completeness. $T_c(q)$ vanishes
abruptly at $q=q_c$.}  \label{Tc-classic}
\end{figure}

The full solution of Eq.(\ref{Tc3}) has three roots, among
which the physical one is confined to the finite $q$-range as
shown by the bold line in Fig.\ref{Tc-classic}. A (physical)
solution exists thus only within the finite range between
clean limit ({\it CL}) and dirty limit ({\it DL}).
One may expand and control this physical solution (of the cubic equation)
in the weak disorder regime (small $q$), where the $T_c$-shift obeys
\begin{equation}
\frac{1}{T_c} = \frac{1}{T_c^{(1)}}+ \frac{1}{\sqrt{\eta
\epsilon_F E_{\perp}}} +
 \frac{1}{T_c^{(1)}} \bigg(\frac{W}{2 E_{\perp}} \bigg)^2,
\label{criticalT}
\end{equation}
showing that $T_c$ decreases with increasing randomness like
$W^{2}$. For a pure system Eq.(\ref{criticalT}) gives the
dependence  $T_c \sim E_{\perp}^{1/2}$, in agreement with Efetov
and Larkin in Ref. \onlinecite{el74}. This expression shows that even a
small interchain-coupling sustains ODLRO in the system and,
consequently, the critical temperature increases with $\sqrt{E_{\perp}}$.
On the other hand, the competing destructive effect of disorder
reduces $T_c$ due to "melting" of the order parameter phase coherence
between neighboring chains.

According to the (physical) solution of Eq.(\ref{Tc3}),
the critical temperature decreases monotonically with
increasing $q$ but finite $T_c$ are confined to the interval
$0 \le q \le q_c =\frac{2}{3 \sqrt 3}$.
The SC-phase becomes fully suppressed for disorder-strengths $W^2$
exceeding a critical disorder-value $W_c^2$ given by
\begin{equation}
W_c^2 =\frac{4 E_{\perp} T_c^{(1)}}{3}
\sqrt{\frac{E_{\perp}}{3 \eta \epsilon_F}},
\end{equation}
beyond which the system is in a normal metallic phase
(for $W^2 > W_c^2$).
The critical temperature drops to zero at $W^2 = W_c^2$ with a
jump of size
\begin{equation}
\Delta T_c = T_c^{\ast}= \left( \sqrt{\frac{3}{\eta \epsilon_F
E_{\perp}}} + \frac{1}{T_c^{(1)}}\right)^{-1}.
\end{equation}
Thus the SC-normal metal phase transition appears as a
$1$st-order transition.

In order to describe the behavior of $t$ near the disorder limit {\it DL}
(see Fig.\ref{Tc-classic}) we expand around
$\{t^{\ast}, q^{\ast}\} = \{\sqrt{3},\frac{2}{3 \sqrt{3}} \}$,
in terms of small (nonnegative) $\delta t = t^{\ast} - t$ and
$\delta q = q^{\ast} - q$, which gives
$\delta t = 3^{3/4}\sqrt{\delta q}$.

Near the dirty limit, the $T_c$-variation has an infinite slope
(see also Fig.1). This can be reexpressed in terms of the physical
parameters $\delta T_c =T_c - T_c^{\ast}$ and the variance $W^2$
of the Josephson-coupling, by reinserting
Eqs.(\ref{q-eq},\ref{criticalT}), as
\begin{equation}
\delta T_c = \frac{3^{3/4} T_c T_c^{\ast}}{E_{\perp}(4 \eta
\epsilon_F E_{\perp}(T_c^{(1)})^2)^{1/4}} (W_c^2 - W^2)^{1/2}
\end{equation}
in the vicinity of the breakdown point
\begin{equation}
\{ T_c^{\ast}, W_c^2 \} = \bigg\{ \left( \sqrt{\frac{3}{\eta
\epsilon_F E_{\perp}}} + \frac{1}{T_c^{(1)}}\right)^{-1}, \frac{4
E_{\perp} T_c^{(1)}}{3} \sqrt{\frac{E_{\perp}}{3 \eta \epsilon_F}}
\bigg \}.
\nonumber
\end{equation}
Thus the critical temperature decreases with disorder almost
linearly but, approaching the dirty limit ${\it DL}$, it finally turns
into a (nonanalytic) square-root behavior close to the breakdown point.
In the absence of the non-magnetic disorder, even arbitrarily small
Josephson-coupling between the chains stabilizes the ODLRO and gives
a non-zero critical temperature. However, the SC phase with
finite Josephson-coupling can be destroyed by increasing the
strength of non-magnetic disorder.
%

\section{Quantum phase fluctuations}
The self-consistent mean-field method, applied above for the
classical phase-fluctuation regime, expressed the $T_c$-equation
in terms of the pure $1$D phase-correlator (\ref{1Dcorrelator}),
neglecting in this respect the Josephson-coupling between
neighboring chains. We shall now improve the calculation of the
phase-correlator by taking into account the transverse rigidity of
the system, which provides a more realistic determination of the
transition temperature in the quantum fluctuation regime. Our
calculations are carried out in the Hamilton formalism for
convenience, yet the problem can be formulated  in the path
integral language \cite{el74} as well.

Let us start from the Lagrangian, again taking ${\bf B}=0$ for
simplicity,
\begin{equation}
\mathcal{L} = \frac{K \xi_{\|}(0)}{8} \sum_{\bf j} \int dz [\hbar
\dot{\varphi}_{\bf j}(z)]^2 - F_{st}^{eff}\{\varphi \},
\label{lagrangian}
\end{equation}
where $\dot{\varphi}$ denotes the time derivative of the phase.
The dynamical term in the Lagrangian can be interpreted as the
electrostatic energy of charged wires \cite{fy76, nf98}
\begin{equation}
E_{el}=\frac{1}{2} \sum_{\bf i,j} \int dz \int dz'\hspace{.1cm}
C_{\bf i,j}(z-z')V_{\bf i}(z)V_{\bf j}(z'),
\label{Elec}
\end{equation}
generated according to the first Josephson equation $\dot{\varphi}
= (2e/\hbar)V$, and $C_{\bf i,j}(z-z')$ are the specific coefficients
of electrostatic induction. Rewriting the electrostatic energy
$E_{el}$ in terms of the time-derivative of phases, the Fourier
transform $K({\bf q_{\perp}}, q_z)$ of  the coefficients
$K_{\bf i,j}(z-z')=\frac{1}{4e^2}C_{\bf i,j}(z-z')$, has the
physical meaning of a compressibility.  In Eq.(\ref{lagrangian})
we neglect dispersion in the compressibility and assume
$K({\bf q_{\perp}},q_z) = K = const.$ This approximation is
equivalent to a replacement of the electrostatic energy
(\ref{Elec}) by $\frac{1}{2}C\int dz \sum_j V_j^2(z)$. The
parameter $K$ can be calculated \cite{el74} in the presence of
Coulomb screening for a small Born parameter $\frac{e^2}{\hbar
v_0} n <1$, which results in
\begin{equation}
K = \frac{n}{\pi \hbar v_0} \left[1 + \frac{e^2}{\pi \hbar v_0} n
\ln \frac{a_{\perp}}{d} \right]^{-1}.
\label{compres}
\end{equation}
There, $K_0 = \frac{n}{\pi \hbar v_0}$ is the unscreened compressibility,
$v_0$ denotes the longitudinal velocity of an electron on the
Fermi surface averaged over $n$ subbands, while $a_{\perp}$ and $d$
stand for the interchain-distance and the diameter
of a superconducting wire, respectively.

In order to get $F_{st}^{eff}\{\varphi \}$ we average the free
energy over disorder and apply a mean-field approximation,
corresponding to the replacement $\int dz \int dz' [1-\cos
(\varphi_{\bf j}(z) - \varphi_{\bf j+g}(z))][1-\cos (\varphi_{\bf
j}(z') - \varphi_{\bf j+g}(z'))]$ by $\langle \langle [1 - \cos
(\varphi_{\bf j}(z) - \varphi_{\bf j+g}(z))]\rangle \rangle \int
dz [1-\cos (\varphi_{\bf j}(z) - \varphi_{\bf j+g}(z))].$

The Hamiltonian, expressed through the phases $\varphi_{\bf j}$
and canonical conjugate momenta $\Pi_{\bf j}$, becomes
\begin{eqnarray}
\mathcal{H} = \sum_{\bf j} \int dz \bigg \{ 2 \frac{ \Pi_{\bf
j}^2(z)}{K \xi_{\|}(0)} + \frac{\hbar^2 N_s^{(1)}(T)}{8m_{\|}
\xi_{\|}}\biggl[ \bigg( \frac{\partial
\varphi_{\bf j}}{\partial z} \bigg)^2 +{}
                      \nonumber\\
 {} + \sum_{\bf g} \delta_{cl}^2 [1 - \cos(\varphi_{\bf j}(z) - \varphi_{\bf j + g}(z))]\biggr]
\bigg\},
\label{hamilton}
\end{eqnarray}
where $\Pi_{\bf j} = \frac{1}{\hbar}\frac{\delta {\mathcal
L}}{\delta \dot{\varphi}_{\bf j}}=\frac{1}{4}\hbar K \xi_{\|}(0)
\dot{\varphi}_{\bf j}$, while $\delta_{cl}$ is given by
\begin{equation}
\delta_{cl}^2 = \delta_{0}^2 \biggl[ 1 - \frac{W^2 N_s^{(1)}
\xi_{\|}}{E_{\perp} T} \langle \langle [1 - \cos (\varphi_{\bf
j}(z) - \varphi_{\bf j+g}(z))]\rangle \rangle \biggr]
\label{rigidity-dis}
\end{equation}
and represents the dimensionless anisotropy-parameter or the
transverse rigidity of the system; $\delta_{0}$ in
Eq.(\ref{rigidity-dis}) is transverse rigidity of the pure
system
\begin{equation}
\delta_{0} = \left(\frac{E_{\perp}}{\hbar^2/8 m_{\|} \xi_{\|}^2}
\right)^{1/2} = \frac{(\epsilon_F E_{\perp})^{1/2}}{T_c^{(1)}}.
\label{rigidity-pure}
\end{equation}

The phase dynamics in the classical limit can be obtained from the
Euler-Lagrange equation, which is described by a set of coupled
sine-Gordon type non-linear equations
\begin{equation}
\ddot{\varphi}_{\bf j}(z) = \bar{\omega}^2 \bigg \{
\frac{\partial^2 \varphi_{\bf j}}{\partial z^2} -  \sum_{\bf g}
\delta_{cl}^2
\sin (\varphi_{\bf j}(z) - \varphi_{\bf j+g}(z)) \bigg \},
\label{dynamics}
\end{equation}
where $\bar{\omega}$ is a characteristic scale of frequency, and
\begin{equation}
\bar{\omega}^2=\frac{N_s^{(1D)}(T)}{m_{\|} K \xi_{\|}^2}.
\end{equation}

We express $\bar{\omega}$ as $\bar{\omega} = 2 \pi \alpha
\tau^{1/2} T_c^{(1)}/\hbar$, where
\begin{equation}
\alpha = \frac{1}{2 \pi}\bigg(\frac{16}{ K (\hbar^2
N_s^{(1D)})/m_{\|}}\bigg)^{1/2}.
\label{alpha}
\end{equation}

The parameter $\alpha$ is the essential parameter of the theory,
which can be written, using Eq.(\ref{compres}), as
\begin{equation}
\alpha =\frac{1}{\sqrt {\pi n}} \left[1 + \frac{e^2}{\pi \hbar
v_0} n \ln \left(\frac{a_{\perp}}{d} \right) \right]^{1/2}.
\end{equation}
The system of equations (\ref{dynamics}) is linearized for small
$\varphi_{\bf j}$ and its Fourier-transformation becomes diagonal
\begin{equation}
\ddot{\varphi}({\bf q_{\perp}},q_z) = -\omega^2({\bf
q_{\perp}},q_z) \varphi ({\bf q_{\perp}},q_z) \label{diagonal}
\end{equation}

The eigenfrequency of oscillations $\omega({\bf q_{\perp}},q_z)$
is given in the harmonic approximation as
\begin{equation}
\omega ({\bf q}, q_z) = \bar{\omega} [q_z^2 + \delta _{cl}^2 2 (2-
\cos q_x - \cos q_y)]^{1/2}. \label{frequency}
\end{equation}
These equations describe the frequency of low-lying plasmon-mode
of the system.

The quantum description is realized by expressing $\varphi_{\bf
q}$ and $\Pi_{\bf q}$ as a linear superposition of Bose operators
$b_{\bf q}$ and $b_{\bf q}^{\dag}$, $[b_{\bf q}, b_{\bf q}^{\dag}] = 1$,
by
\begin{eqnarray}
&&\varphi_{\bf q} = \left ( \frac{\pi \alpha {\bar
\omega}}{\omega ({\bf q})} \right )^{1/2} (b_{\bf q} + b_{- \bf
q}^{\dag})\nonumber\\
&&\Pi_{\bf q} = i \left( \frac{\omega ({\bf q})}{4
\pi \alpha \bar{\omega}} \right)^{1/2} (b_{- \bf q} - b_{\bf
q}^{\dag}).
\end{eqnarray}
If we expand the cosine term in Eq.(\ref{hamilton})
up to the quadratic term and express the phases $\varphi_{\bf q}$
and the conjugate momentum $\Pi_{\bf q}$ in terms of creation and
annihilation operators, we get the Hamiltonian in the harmonic
approximation as
\begin{equation}
\hat{\mathcal H}_0 = \sum_{\bf q} \hbar \omega ({\bf q}_{\perp},
q_z) \left[ b_{\bf q}^{\dag} b_{\bf q} + \frac{1}{2} \right],
\label{hamiltonian}
\end{equation}
where the energy spectrum is defined by Eq.(\ref{frequency}).

In order to take into account the quantum effects in the
Hamiltonian, we have to express the cosine term in
Eq.(\ref{hamilton}) in a normal ordering before expanding over
$b_{\bf q}^{\dag}$ and $b_{\bf q}$
\begin{eqnarray}
&& 1 - \cos (\varphi_{\bf j} (z) - \varphi_{\bf j + g}(z)) =
\nonumber\\
&&\hspace{-1cm}1-\frac12 e^{- S_{\alpha}({\bf g}, 0)}\left( e^{\sum_{\bf q}
A_{\bf q} b_{\bf q}^{\dag}} e^{- \sum_{\bf q} A_{\bf q}^{\star}
b_{\bf q}} +
h.c.\right),
\label{operator}
\end{eqnarray}
where
\begin{equation}
A_{\bf q}({\bf j}, {\bf g}) = i \bigg( \frac{\pi \alpha
\bar{\omega}}{N \omega ({\bf q})} \bigg)^{1/2} e^{i q_z z + i {\bf
q_{\perp} j}} (1 - e^{i {\bf q_{\perp} g}} ).
\label{coeff}
\end{equation}

The prefactor $\exp \{- S_{\alpha}({\bf g},
0)\}$ originates in the commutation relation between $b_{\bf
q}^{\dag}$ and $b_{\bf q}$, and by taking into account the
Baker-Campbell-Hausdorff relation
$\exp ({\hat H}_1 + {\hat H}_2 ) = \exp
({\hat H}_1) \exp ({\hat H}_2) \exp \{ -\frac{1}{2} [{\hat H}_1,
{\hat H}_2 ] \}$. Furthermore
\begin{equation}
S_{\alpha}({\bf g},0) = \frac{1}{2} \sum_{\bf q} |A_{\bf q}({\bf
j, g})|^2 = \frac{\pi \alpha \bar{\omega}}{N} \sum_{\bf q} \frac{1
- \cos ({\bf q}_{\perp} {\bf g})}{\omega ({\bf q}_{\perp},q_z)},
\label{S}
\end{equation}
where $N$ is the number of unit cells per volume.

It is clear from Eq.(\ref{operator}) that the physical meaning of
$\exp \{- S_{\alpha}({\bf g}, 0)\}$ is an average of
$\cos(\varphi_{\bf j}(z) - \varphi_{\bf j+g}(z))$ over the ground
state at $T = 0$. In the framework of the self-consistent phonon
approximation (SCPA), we expand the Josephson term (\ref{operator})
in powers of the creation  and annihilation operators,
$b_{\bf q}^{\dag}$ and $b_{\bf q}$ repsectively.
Expressing the leading (harmonic) part of the Hamiltonian (\ref{hamilton})
in terms of the particle number operator
${\hat N}_{\bf q} = b_{\bf q}^{\dag} b_{\bf q}$,
we obtain again a harmonic Hamiltonian as in Eq.
(\ref{hamiltonian}), $\hat{\mathcal H}_0 \to \hat{\mathcal
H}_{\alpha}^{(0)}(0)$, but with the different oscillation frequency
\begin{equation}
\omega ({\bf q}, q_z) = \bar{\omega}[q_z^2 + \delta _{cl}^2 e^{-
S_{\alpha}^{(0)}({\bf g}, 0)} 2 (2- \cos q_x - \cos q_y)]^{1/2},
\label{frequency0}
\end{equation}
Thus the application of the SCPA results in a
re-normalization of the parameter of anisotropy $\delta_{cl}$ in
the frequency of oscillation $\omega ({\bf q}, 0)$ by means of the
phase-phase correlator as
\begin{equation}
\delta_{cl}^2 \to \delta_{qu}^2 (0) = \delta_{cl}^2
e^{-S_{\alpha}({\bf g}, 0)}.
\label{delta0}
\end{equation}
We note that in order to preserve a maximal coherence of phases
at $\varphi_{\bf j} = \varphi_{\bf j+g}$ in the Josephson term
$1-\cos({\hat\varphi}_{\bf j}(z)-{\hat\varphi}_{\bf j+g}(z))$
the latter is replaced in the framework
of the SCPA by $e^{-S_{\alpha}({\bf g}, 0)} - \cos ({\hat
\varphi}_{\bf j}(z) - {\hat \varphi}_{\bf j + g}(z))$, which
corresponds to shifting of the energy origin. Indeed, the
Josephson term in the initial expression of the Hamilton function
was introduced in a such way that it becomes zero for a maximal
coherence of phases $\varphi_{\bf j} = \varphi_{\bf j+g}$.
Zero-point fluctuations at $T=0$ in the quantum case destroy the
phase coherence and increase the Josephson energy. By shifting $1
\to e^{- S_{\alpha}({\bf g},0)}$ we again reach a minimal
Josephson energy at $T=0$ in the quantum case too. Similar
shifting was done also in Eq.(\ref{rigidity-dis}).

In the expansion of the exponential operator of Eq.(\ref{operator})
we select all diagonal terms, which can be expressed in terms
of the bosonic particle number operator
${\hat N}_{\bf q} = b_{\bf q}^{\dag} b_{\bf q}$  \cite{nf98}.
This yields
\begin{eqnarray}
&&\{ \cos (\varphi_{\bf j}(z) - \varphi_{\bf j+g}(z)) \}_{diag} =
e^{- S_{\alpha}({\bf g},0)} \prod_{\bf q} \bigg \{1 -
\nonumber\\
&&\hspace{-.9cm}- |A_{\bf q}|^2 b_{\bf q}^{\dag} b_{\bf q} + |A_{\bf q}|^4
\frac{(b_{\bf q}^{\dag})^2}{2 !} \frac{b_{\bf q}^2}{2 !} - |A_{\bf
q}|^6 \frac{(b_{\bf q}^{\dag})^3}{3 !} \frac{b_{\bf q}^3}{3 !} \pm
\dots  \bigg \}.
\label{expeqn}
\end{eqnarray}
Absorbing now the product in Eq.(\ref{expeqn}) into the exponential form,
and neglecting all higher orders in $|A_{\bf q}|^2$ beyond the
leading term $|A_{\bf q}|^2$ (for justification see below
[\onlinecite{phonon}]),
we get
\begin{eqnarray}
\{ \cos (\hat{\varphi}_{\bf j} - \hat{\varphi}_{\bf j+g})
\}_{diag} = \exp \{-S_{\alpha}({\bf g},0) - \sum_{\bf q} |A_{\bf
q}|^2 {\hat N}_{\bf q} \} =
\nonumber\\
e^{-S_{\alpha}({\bf g}, T)} \equiv \exp \left\{-\frac{2 \pi \alpha
\bar{\omega}}{N} \sum_{\bf q} \frac{1 - \cos ({\bf q}_{\perp} {\bf
g})}{\omega({\bf q})} \left({\hat N}_{\bf q} +
\frac{1}{2}\right)\right\}. \nonumber
\end{eqnarray}

Thus, after this step of calculations we still restrict ourself with
harmonic approximation, describing the system by the Hamiltonian
$\hat{\mathcal H}_{\alpha}^{(0)}(T)$ like (\ref{hamiltonian})
where the transverse rigidity $\delta_{cl}^2$ in the frequency
$\omega ({\bf q}, T)$ is renormalized as
\begin{equation}
\delta_{cl}^2 \to \delta_{qu}^2(T) = \delta_{cl}^2 \exp\{-
S_{\alpha}({\bf g},T) \},
\label{delta-quT}
\end{equation}
where
\begin{eqnarray}
\langle \langle \cos (\varphi_{\bf j}(z) - \varphi_{\bf
j+g}(z))\rangle \rangle \equiv e^{-S_{\alpha}({\bf g},T)}
\nonumber\\
=\frac{ Tr \big \{ e^{-\beta {\hat H}_{\alpha}^{(0)}} \cos({\hat
\varphi}_{\bf j}(z) - {\hat \varphi}_{\bf j+g}(z)) \big \}}{ Tr
\big \{ e^{-\beta \hat{\mathcal H}_{\alpha}^{(0)}} \big \}}.
\label{corrT}
\end{eqnarray}

The trace over the diagonal part of the phase-phase correlator
within the harmonic approximation replaces the bosonic filling
number operator ${\hat N}_{\bf q}$ by the Planck distribution
function for phonons with energies of $\hbar \omega ({\bf q},T)$
as ${\hat N}_{\bf q} \to \{ \exp \left( \frac{\hbar \omega ({\bf
q},T)}{T} \right) - 1\}^{-1}$, yielding the following expression
for the $S_{\alpha}({\bf g},T)$
\begin{equation}
S_{\alpha}({\bf g},T) =\frac{\pi \alpha \bar{\omega}}{N} \sum_{\bf
q} \frac{1 - \cos ({\bf q}_{\perp} {\bf g})}{\omega ({\bf q}, T)}
\coth \left( \frac{\hbar \omega ({\bf q},T)}{T} \right).
\label{correlatorT}
\end{equation}

The correlator $e^{-S_{\alpha}({\bf g},T)}$ and its $T=0$-limit,
as given by Eqs.(\ref{correlatorT}) and (\ref{S}) respectively,
are evaluated explicitly in the Appendix.
\subsection{Quantum Criticality at T=0}
The zero-temperature behavior of the system is analyzed by means of
the phase-phase correlator $e^{-S_{\alpha}({\bf g},0)}$, the explicit
expression for which is given by Eq.(\ref{S(g,0)3}) in the Appendix.
Expressing the phase-phase correlator $e^{-S_{\alpha}({\bf g},0)}$
in terms of $S_{\alpha}({\bf g},0)$ gives $e^{-S_{\alpha}({\bf
g},0)} = (\delta_{qu}(0))^{\alpha}\equiv \delta_{qu}^{\alpha}$,
which implies that even a small interchain-coupling stabilizes ODLRO
in the system, hence also a finite $T$ phase transition should
exist. In order to get an explicit expression for the dependence
of $ \delta_{qu}$ on $\delta_0$ and on disorder, we have to solve
the equation $\delta_{qu}^2 = \delta_{cl}^2 e^{-S_{\alpha}({\bf
g}, 0)}$ together with Eq.(\ref{rigidity-dis}) for $\delta_{cl}$.
Thus the equation for the reduced transverse rigidity
$\delta_{qu}^{\ast} = \delta_{qu}/\delta_{qu}^{(0)}$,
where $\delta_{qu}^{(0)} = \delta_0^{\frac{2}{2 - \alpha}}$
is the renormalized transverse rigidity for the clean system
at $T = 0$, assumes the form
\begin{equation}
(\delta_{qu}^{\ast})^{3 - 2\alpha} = (\delta_{qu}^{\ast})^{1 -
\alpha} - q_{qu},
\label{delta-qu}
\end{equation}
where the quantum parameter of randomness $q_{qu}$ reads
\begin{equation}
q_{qu} = \frac{C W^2}{2 E_{\perp}^2} \delta_0^{\frac{2}{2 -
\alpha}}.
\label{qofW}
\end{equation}
$C$ in Eq.(\ref{qofW}) is a constant $C \sim 1$. Although
Eqs.(\ref{Tc3}) and (\ref{delta-qu}) are written for two different
characteristic parameters of the system, it is easy to see that the
equation for $\delta_{qu}^{\ast}(0)$, if we neglect the quantum
effects at $\alpha = 0$, coincides with Eq.(\ref{Tc3}) written for
$y=1/t$.

Eq.(\ref{delta-qu}) can be solved approximately for moderately
weak disorder, yielding the following expression for
$\delta_{qu}(0)$
\begin{equation}
\delta_{qu}(0) = \delta_0^{\frac{1}{1-\alpha /2}}\left[1 - C
\frac{1 -\alpha}{2 - \alpha} \left( \frac{W}{E_{\perp}} \right)^2
\delta_0^{\frac{1}{1 - \alpha/2}} \right]^{\frac{1}{1 - \alpha}}.
\label{delta-qu0}
\end{equation}
Hence, the evaluation of the phase-phase correlator $\langle \langle \cos
(\varphi_{\bf j}(z) - \varphi_{\bf j+g}(z))\rangle \rangle|_{T=0} =
e^{-S_{\alpha}({\bf g},0)}$ in the presence of disorder
yields the result
\begin{eqnarray}
&&\langle \langle \cos (\varphi_{\bf j}(z) - \varphi_{\bf
j+g}(z))\rangle \rangle|_{T=0} =
\left(\frac{\epsilon_F E_{\perp}}{{T_c^{(1)}}^2}
\right)^{\frac{\alpha}{2 - \alpha}}\times
\nonumber\\
&&\times \left[1 - C \frac{1 -\alpha}{2 - \alpha} \left(
\frac{W}{E_{\perp}} \right)^2 \left(\frac{\epsilon_F
E_{\perp}}{{T_c^{(1)}}^2}\right)^{\frac{1}{2 - \alpha}}
\right]^{\frac{\alpha}{1 - \alpha}}.
\label{S(g,0)F}
\end{eqnarray}
In the absence of the disorder, i.e. for $W = 0$, we retrieve the
phase-phase correlator
\begin{equation}
\langle\cos(\varphi_{\bf j}(z)-\varphi_{\bf j+g}(z))\rangle|_{T=0,W=0}
= \left(\frac{\epsilon_F E_{\perp}}{{T_c^{(1)}}^2}
\right)^{\frac{\alpha}{2 - \alpha}},
\nonumber\\
\end{equation}
as obtained by Efetov and Larkin \cite{el74} for pure quasi-1D
superconductors. The critical temperature for a quasi-1D
superconductor, according to Efetov and Larkin \cite{el74}, is
defined by $T_{c0} = \delta_{qu}^{(0)}(0) T_c^{(1)}$ for pure
superconductors and by $T_{c0} = \delta_{qu}(0) T_c^{(1)}$ in our
case for dirty superconductors. This relation with
Eq.(\ref{S(g,0)F}) shows that $T_{c0}$ decreases nonlinearly with
disorder.
\begin{figure}
\hspace{-.5cm}\resizebox{.5\textwidth}{!}{%
\includegraphics[width=1cm]{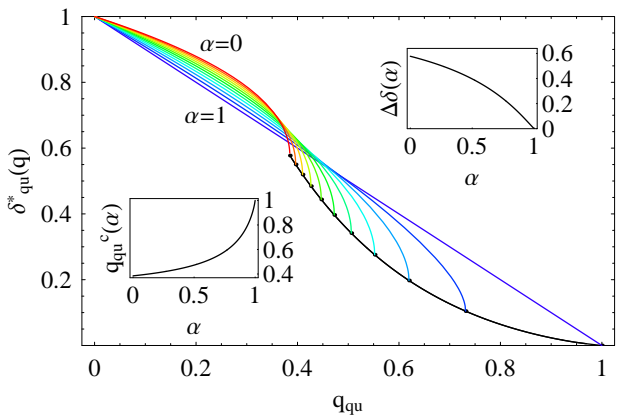}}
\caption{The dependence of the reduced $T=0$ transverse rigidity
$\delta_{qu}^{\ast}(q_{qu})$ on the disorder-strength parameter
$q_{qu}$ is shown for $0\leq\alpha\leq 1$ in steps of
$\Delta\alpha=0.1$. At $q_{qu}=q_{qu}^c$,
$\delta_{qu}^{\ast}(q_{qu})$ drops to zero for $\alpha <1$ and
vanishes continuously only at $\alpha =1$. Inserts show the
$\alpha$-variation of the jump (upper right corner) and of its
position $q_{qu}^c(\alpha)$ (lower left).} \label{Tc-quantum}
\end{figure}

The numerical solution of Eq.(\ref{delta-qu}) is depicted in
Fig.\ref{Tc-quantum}. The reduced $T=0$ transverse rigidity
$\delta_{qu}^{\ast}(q_{qu})|_{T=0}$ is shown to decrease with
increasing disorder for (fixed) $\alpha < 1$, and suddenly drops
to zero at the critical disorder strength $q_{qu} = q_{qu}^{c}$.
Hence the quantum critical behavior corresponds to a first order
phase transition. Fig.\ref{Tc-quantum} shows how the breakdown
point shifts with increasing $\alpha$ to higher values of the
randomness, and the jump vanishes as $\alpha\rightarrow 1$.
Eq.(\ref{delta-qu}) becomes linear for $\alpha = 1$ and gives, by
inferring the $q_{qu}(W)$-relation from Eq.(\ref{qofW}),
\begin{equation}
\delta_{qu}(W)|_{T=0,\alpha=1} = \delta_0^2 \left[ 1 - \frac{C W^2
\epsilon_F}{2 E_{\perp} \left(T_c^{(1)}\right)^2}\right].
\end{equation}
Here, the transverse rigidity $\delta_{qu}(W)|_{T=0,\alpha=1}$
decreases linearly with increasing $W^2$ and vanishes at
\begin{equation}
W_c^2 =
\frac{2 E_{\perp} \left(T_c^{(1)}\right)^2}{C \epsilon_F}.
\end{equation}
The quantum critical behavior in the model is however controlled
by two parameters, the strength of randomness $q_{qu}(W)$ and the
parameter of quantum dynamics $\alpha$. For $\alpha < 1$, the
superconductor-normal metal phase transition at $T=0$ is always
discontinuous, and only turns into second-order at $\alpha = 1$.
\subsection{Phase transition at finite temperatures}
Let us now study the {\it finite $T$ behavior} of the transverse
rigidity. The phase transition in a quasi-1D superconductor occurs
at some temperature $T = T_c$ when the transverse rigidity in the
ensemble of phases $\{ \varphi_{\bf j}(z) \}$ vanishes, which
results in "melting" of the phase coherence. The energy spectrum
$\omega ({\bf q}_{\perp}, q_z)$ of the collective excitations is
reorganized and the transverse ${\bf q}_{\perp}$-dependent part of
$\omega ({\bf q}_{\perp}, q_z)$ vanishes at $T = T_c$, i.e.
symmetry breaking occurs in the bosonic excitation at $T = T_c$.
Inserting the solution of Eq.(\ref{S}) for $T < \alpha T_c^{(1)}$
into $\delta_{qu}^2(T) = \delta_{cl}^2 e^{- S_{\alpha}({\bf g},
T)}$ and using Eq.(\ref{delta-quT}), we obtain
\begin{equation}
\delta_{qu}^2(T) = \delta_{qu}^2(0)\left(\frac{T}{\alpha
T_{c0}}\right)^{\alpha} \exp\bigg \{-C
\frac{T}{T_{c0}}\frac{\delta_{qu}(0)}{\delta_{qu}(T)}\bigg \},
\label{deltaTC}
\end{equation}
where a new temperature scale is introduced by means of $T_{c0} =
\delta_{qu}(0) T_c^{(1)}$. In terms of $y =\left(\frac{\alpha
T_{c0}}{T}\right)^{\alpha /2}
\frac{\delta_{qu}(T)}{\delta_{qu}(0)}$ and $\theta = \left(
\frac{T}{T_{c0}} \right)^{1 -
\frac{\alpha}{2}}\frac{C}{2}\alpha^{\alpha/2}$, Eq.(\ref{deltaTC})
assumes the form $y = exp \{ - \theta /y \}$, which has a non-zero
solution only for $\theta \leq e^{-1}$. The finite solution of
this equation vanishes discontinuously at $\theta = \theta_c=
e^{-1}$, giving the following value for $T_c$
\begin{equation}
T_c = \delta_{qu}(0)T_c^{(1)}{\alpha}^{- \frac{\alpha}{2 -
\alpha}} (2/e C)^{\frac{2}{2 - \alpha}}.
\label{fig3-curve}
\end{equation}
\begin{figure}
\hspace{-.5cm}\resizebox{.47\textwidth}{!}{%
\includegraphics[width=1cm]{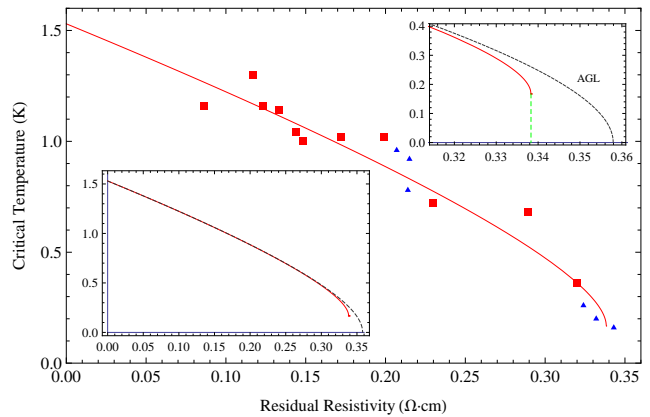}}
\caption{Comparison of the $T_c$-decrease, as obtained from
Eq.(\ref{fig3-curve}) with fit parameter $\alpha=0.47$ of the
present theory (after proper rescaling), with experimental data
imported from Ref.\onlinecite{japj05} by Joo et al. Different
symbols belong to different samples and different slow (squares)
or fast (triangles) cooling procedures. In a first approximation
the theoretical solid curve, as shown from the clean limit up to the
dirty limit, fits and confirms the slightly but increasingly
non-linear behavior with a final discontinuous drop of $T_c$ at
the critical disorder. In the lower left inset $T_c$-curve employing
the Abrikosov-Gor'kov-Larkin (AGL) digamma-function (dashed curve)
for unconventional superconductors is shown for comparison with the
present theory: the AGL-curve is chosen such that its value and
slope at zero disorder agrees with the present one.
The insert in the upper right corner shows the remarkable deviation
of these two theoretical curves in the large disorder regime
and close to the breakdown.}
\label{exper}
\end{figure}
The magnitude of the jump in $y(\theta_c)$ is $e^{-1}$, and hence
the phase transition is of first-order. The similar behavior has
been found also in the planar rotor model \cite{pu73} in the
absence of disorder. The dependence of $T_c$ on the disorder is
determined by the zero-temperature transverse rigidity
$\delta_{qu}(0)$, the behavior of which is depicted in Fig.
\ref{Tc-quantum}. Therefore, for arbitrary $\alpha < 1$ the
critical temperature decreases monotonically with increasing
randomness and drops to zero at the critical disorder strength
$q_{qu} = q_{qu}^c$.
The variation of the critical temperature versus the residual
resistivity of the organic superconductor
$(TMTSF)_2(ClO_4)_{(1-x)}(ReO_4)_x$ has been explored by Joo et
al. in Ref.\onlinecite{japm04,japj05}.  The experimental data are
read off from Ref.\onlinecite{japj05} and copied into
Fig.\ref{exper} in order to provide a close comparison with a
theoretical fit-curve as obtained from Eq.(\ref{fig3-curve}) of
the present theory.

In this first approach and within a moderate accuracy, the data as
published in the paper of Joo et al., eventually appear to find an
explanation by our theory. In order to establish the link between
experiment and theory, the following argument is exploited: the
substitution of $ClO_4$ anions by $ReO_4^-$ in the relaxed (R)
samples increases the residual resistivity, which is proportional
to the inverse lifetime $1/\tau$ of electron as well as to the
disorder strength $W^2$ (as used in the present theory).

For the relaxed samples of Ref.\onlinecite{japj05}, referred to as
the R-sample(s), the doping concentration $x$ varies in the
interval of $0 \le x \le 0.1$ under slow cooling from the clean
sample to the nominal concentration $x=0.1$. Slow cooling of
$R$-samples assures a uniform orientation of the anions along the
stacking axis, whereas fast cooling in the (quenched) $Q$-samples
introduces strong orientational disorder and increases the
residual resistivity. $T_c$ decreases quasi-linearly with
increasing disorder (or the residual resistivity) in the large
interval of the randomness. Around the breakdown point the
dependence of $T_c$ on randomness is non-linear. The critical
doping concentration, corresponding to the breakdown of the
superconducting state, grows with the quantum charging effect in
the system. Substitution of $ClO_4$ anions by $ReO_4$ seems to
increase the quantum charging parameter $\alpha$, shifting thus
the breakdown point to a higher value of the residual resistivity.
All these features and experimental evidences agree well with the
theory.

In order to compare with other well-known cases of $T_c$-suppression
by disorder we consider the pair breaking theory for a
superconductor with unconventional gap symmetry
\cite{larkin65,ycwlm03}. This physically different case of
$T_c$-reduction by {\it non-magnetic} impurities in {\it unconventional}
superconductors was found to be described by the famous
digamma-formula of Abrikosov-Gor'kov's conventional pair breaking
theory \cite{ag61} in the presence of {\it paramagnetic} impurities.
The $T_c$-reduction may thus be expressed in the form
\begin{equation}
\ln \left(\frac{e^{\Psi(\frac12)}T_c^{(1)}}{T_c}\right) = \Psi
\left(\frac{1}{2}+\frac{\rho T_c^{(1)}}{2\pi T_c}\right),
\label{digamma}
\end{equation}
where $\Psi$ means the digamma function, $\rho = \hbar/2\tau
T_c^{(1)}$ is the depairing parameter, and $\tau$ the elastic
scattering time. Both theoretical curves, as shown in the insets
of Fig.\ref{exper}, can be chosen to coincide for weak and
moderately strong disorder (where the linear decay is rather
unspecific). Approaching the SC-breakdown at larger disorder they
differ however substantially. The AGL-solution for unconventional
pairing approaches $T_c=0$ continuously and obeys a square root
dependence $T_c\sim (\tilde{q}-\tilde{q}_c)^{1/2}$, where
$\tilde{q}$ stands for the disorder-strength in the AGL-case. This
square root law follows from the leading ($O(z^2)$) correction of
the digamma-function $\Psi(\frac12+\frac{\tilde q}{z})$ given by
the Laurent series of its exponential
\begin{equation}
\exp\left[\Psi\left(\frac12+\frac{\tilde{q}}{z}\right)\right]=
\frac{\tilde{q}}{z}+\frac{z}{24\tilde{q}}+O(z^2)
\end{equation}
near the logarithmic branch point of $\Psi$ at infinity, hence
$z=0$. In the physical context the variable $z$ corresponds to the
critical temperature $T_c$ of Eq.(\ref{digamma}). By comparing the
exponential of Eq.(\ref{digamma}) one can see that $z=T_c=0$ is
reached for $\tilde{q}=\rho T_c^{(1)}/(2\pi)\rightarrow
\tilde{q}_c=\exp(\Psi(\frac12))$.

By contrast, the present theory does not allow for a continuous breakdown
of superconductivity. According to Fig.3 the suppression of $T_c$
is stronger and an abrupt breakdown occurs at $T_c|_{min}>0$. The
numerical data show a square root behavior however near the
minimal finite $T_c$.

The breakdown point in the curve, corresponding to our theory, seems
to allow for the existence of an intermediate phase, perhaps a glassy
phase below a tricritical point.
\section{Meissner effect}
The current density is calculated according to $\frac{1}{c}{\bf
J}(z,{\bf j}) = - T\frac{\delta}{\delta {\bf A}} \langle \ln
Z({\bf A})\rangle $, where $Z=\int \mathcal{D} \varphi
e^{-F_{st}/T}$. The complete expression for ${\bf J}$ in the
linear response approximation can be obtained after averaging of
$\ln Z$ over disorder (\ref{gauss}) by using Eq.(\ref{freeenergy})
for $F_{st}$. One obtains
\begin{equation}
J_z = \langle \langle \frac{e \hbar N_s^{(1)}}{2 m_{\|}}
\frac{\partial \varphi_{\bf j}}{\partial z} - \frac{e^2 N_s^{(1)}
\xi_{\|}}{m_{\|} c} A_z \rangle \rangle
\end{equation}
for the longitudinal component of the current and
\begin{eqnarray}
{\bf J}_{\perp} = \sum_{\bf g} \frac{2 e a_{\perp} \xi_{\|}
N_s^{(1)}}{\hbar} {\bf g} E_{\bf g} \langle\langle
\sin(\varphi_{\bf j} - \varphi_{\bf j+g})\rangle \rangle -
\nonumber\\
\sum_{\bf g} \frac{4 e^2 a_{\perp}^2 \xi_{\|} N_s^{(1)}}{\hbar^2
c} {\bf g} E_{\bf g} \langle\langle \cos(\varphi_{\bf j} -
\varphi_{\bf j+g})\rangle \rangle ({\bf g}{\bf A}_{\perp})
\end{eqnarray}
for the transverse component of the current.

For simplicity we present here only the diamagnetic contribution
to the $i$-th ($i = \|, \perp $) component of the current
\begin{equation}
J_i^{dia}(z,{\bf j})= - \frac{c}{4\pi \lambda_i^2} A_i(z,{\bf j}),
\end{equation}
where the longitudinal ($\lambda_{\|}$) and the transverse
component ($\lambda_{\perp}$) of the penetration depth are
obtained as
\begin{equation}
\lambda_{\|}^{-2} = \frac{4 \pi e^2 N_s^{(1)}(T)}{c^2 m_{\|}
a_{\perp}^2},
\end{equation}
and
\begin{equation}
\lambda_{\perp}^{-2} = \frac{8 \pi e^2 N_s^{(1)}(T) E_{\perp}}{c^2
\hbar^2}\langle \langle \cos (\varphi_{\bf j} - \varphi_{\bf j+g})
\rangle \rangle .
\end{equation}
While $\lambda_{\|}(T)$ diverges at $T=T_c^{(1)}$ due to pair
breaking in the SC wires, $\lambda_{\perp}(T)$ diverges at the
global SC transition temperature $T=T_c$, where the phase
coherence between neighboring wires is destroyed. The temperature
and the randomness dependencies of $\lambda_{\|}(T)$ and
$\lambda_{\perp}(T)$ also strongly differ each other. The
transverse component of the penetration depth is determined by the
phase-phase correlator, revealing non-linear temperature
dependence and discontinuous behavior at the critical disorder
strength. Nevertheless the longitudinal component of the
penetration depth is given by the conventional London expression
and does not depend on the disorder strength. Randomness in the
Josephson coupling shifts $T_c$ to lower temperatures and,
therefore, the magnetic field parallel to the SC wires penetrates
easier into the organic superconductor. On the other hand, the
type of disorder considered in this article does not break the
Cooper pairs, keeping thus the penetration of a perpendicular
magnetic field into the SC wires unchanged.

\section{Conclusions}
In this paper we report disorder-effects on $T_c$ and on the
diamagnetism of quasi-1D superconductors with random
Josephson-couplings. Interplay of non-magnetic disorder with
quantum phase fluctuations plays a central role for the
superconductor normal-metal phase transitions in this class of
quasi-1D superconductors. Recent experimental data found in
Refs.\onlinecite{japm04, japj05} are shown to be consistent with
the present theory. Quantum criticality is controlled by two
quantities, namely disorder strength and a dynamical parameter of
phase fluctuations. The present model's quantum criticality
signals the existence of a quantum critical phase between SC- and
normal phase. Its nature deserves further investigation.

In our study we neglect the effects of non-linear excitations,
which are a subject of current interest in low dimensional
systems. Note that this topic was explicitly studied by us for
quasi-2D Josephson coupled superconductors in Ref.
\onlinecite{nf98}. As we have shown in the previous section, the
classical motion of the phase is described by a system of coupled
sine-Gordon type non-linear equations (\ref{dynamics}), which
contains non-linear dynamic excitations as well as static
topological defects. The self-consistent phonon approximation
allows us to calculate the phase-phase correlator between two
arbitrary points ${\bf r}=\{z, {\bf j}\}$ and ${\bf r'}=\{z', {\bf
j+g}\}$
\begin{equation}
\langle \langle \cos (\varphi_{\bf j}(z) - \varphi_{\bf
j+g}(z'))\rangle \rangle \equiv e^{-S_{\alpha}(z-z',{\bf g},T)},
\nonumber
\end{equation}
which can be shown to decrease at ${\bf g} \to 0$ and $|z-z'| \to
\infty$ as a power law $\sim (\xi_{\|}/|z-z'|)^{\beta (T)}$,
setting up a quasi-long-range order and implying the existence of
a Berezinskii-Kosterlitz-Thouless (BKT) topological phase
transition \cite{minnhagen87}(perhaps at $T=0$) in a single SC
wire. The critical index $\beta(T)$ contains both phonon and
vortex contributions. Although the phonon contribution to
$\beta(T)$ can be calculated within the SCHA, it is not clear how
the vortex contribution changes the former one. In our knowledge,
the mechanism of excitations of the vortices with opposite
fugacities and their binding in quasi-1D superconductors has not
been adequately studied, and the topic needs further
investigations.

\section{Acknowledgment}
This research was supported by the DFG under grant Op28/7-1.
E.N. thanks D. Jerome for helpful communication. R.O. is grateful
for hospitality and support extended to him by the IPhT at CEA Saclay.

\section{Appendix}
In order to calculate $S_{\alpha}({\bf g},0)$ for ${\bf g} = {\bf
e}_x N_x + {\bf e}_y N_y$ (where ${\bf e}_x, {\bf e}_y$ are unit
vectors and $N_x,N_y$ are the number of unit cells in directions
of $x, y$, correspondingly) we rewrite Eq.(\ref{S}) in the
following form:
\begin{equation}
S_{\alpha}({\bf g},0) = \pi \alpha \int_{-1}^{1} \frac{d q_z}{2
\pi} \int_{- \pi}^{\pi} \frac{d q_x}{2 \pi} \int_{- \pi}^{\pi}
\frac{d q_y}{2 \pi} \frac{1 - \cos ({\bf q}_{\perp} {\bf
g})}{\sqrt{q_z^2 + \delta_{qu}^2(0)\omega_{\perp}^2}},
\label{S(g,0)1}
\end{equation}
where $\omega_{\perp}^2(q_x, q_y) = 2 (2 - \cos q_x - \cos q_y)$
and ${\bf q}_{\perp} = \{q_x, q_y \}$. Introducing a new variable
$z = q_z / \delta_{qu}(0)$ and using the transformation
\begin{equation}
\frac{1}{\sqrt{z^2 + \omega_{\perp}^2}} = \frac{2}{\sqrt{\pi}}
\int_0^{\infty} d t e^{-t^2 (z^2 + \omega_{\perp}^2)},
\label{transform}
\end{equation}
one can integrate out $z, q_x$ and $q_y$ in Eq.(\ref{S(g,0)1}).
Finally, $S_{\alpha} ({\bf g}, 0)$ is expressed as an integral
over $u = 2 t^2$ as
\begin{eqnarray}
&&S_{\alpha} ({\bf g},0)=\frac{\alpha}{2}\int_0^{\infty}\frac{d u}{u}
e^{-2 u} [I_0^2(u) - I_{N_x}(u) I_{N_y}(u)] \times
\nonumber\\
&&\hspace{3.cm}\times \mathrm{erf} \left(\delta_{qu}^{-1}(0) \sqrt{u / 2}\right),
\label{S(g,0)2}
\end{eqnarray}
where $I_N(u)$ is the Bessel function of an imaginary argument,
and $\mathrm{erf}(z) = \frac{2}{\sqrt{\pi}} \int_0^z dt e^{-t^2}$
is the error function. By using the following asymptotic
expressions for $I_N(u)$,
\begin{displaymath}
I_{\nu} (z) = \left\{ \begin{array}{ll} \sum_{k=0}^{\infty}
\frac{1}{k! \Gamma (\nu + k  + 1)}\left(\frac{z}{2} \right)^{\nu +
2k}, &  0 < z < \sqrt{\nu + 1}\\
\frac{e^z}{\sqrt{2 \pi z}}\big[1 - \frac{(\nu + 1/2)(\nu -1/2)}{2
z} \big], & z >> \nu
\end{array} \right.
\end{displaymath}
and for $\mathrm{erf} (z)$,
\begin{displaymath}
\textrm{erf} (z) = \left\{ \begin{array}{ll} \frac{2}{\sqrt {\pi}}
\sum_{k=1}^{\infty} (-1)^{k+1} \frac{z^{2k-1}}{(2k-1) (k-1)!}, & z
< 1\\
1 - \frac{e^{-z^2}}{\sqrt{\pi} z} \sum_{k=0}^{\infty} (- 1)^k
\frac{(2k - 1)!!}{(2 x^2)^k}, & z >> 1
\end{array} \right.
\end{displaymath}
we get for, e.g. $S_{\alpha}(N,0)$ at $N_x = N, N_y = 0$ the
following expression
\begin{equation}
S_{\alpha}(N,0) = \alpha \ln \frac{C_1}{\delta_{qu}(0)} -
\frac{\alpha}{4 \pi} \frac{C_2}{N}, \label{S(g,0)3}
\end{equation}
i.e. $e^{-S_{\alpha}(N,0)} \sim (\delta_{qu}(0))^{\alpha} \exp
\left(\frac{\alpha}{4 \pi} \frac{C_2}{N}\right), \label{S(g,0)4}$
where $C_1$ and $C_2$ are constants of order of unity. If we
take only the first terms in the expansions of $I_N(z)$ and
$\mathrm{erf}(z)$, we get $C_1 = 1/{\sqrt 2}$ and $C_2 = 1$.
Higher order contributions correct only these constants. Thus it is
seen from Eq.(\ref{S(g,0)3}) that the phase-phase correlator in the
transverse direction saturates at $T=0$ to its asymptotic value of
$\delta_{qu}^{\alpha}(0)$ for distances of the unit cell size
$a_{\perp}$.

The correlator $S_{\alpha}({\bf g}, T)$ at $T \neq 0$ is
also calculated in the same way as $S_{\alpha}({\bf g}, 0)$ was
obtained above. Using in Eq.(\ref{correlatorT})
the representation $\coth (\pi z) = \frac{z}{\pi} \sum_{n = -
\infty}^{\infty} \frac{1}{n^2 + z^2}$, the correlator
$S_{\alpha}({\bf g}, T)$ assumes the following form
\begin{eqnarray}
&&S_{\alpha}({\bf g},T)=\frac{T}{\pi \delta_{qu}(T) T_c^{(1)}}
\sum_{n = - \infty}^{\infty} \int_0^{\frac{1}{\delta_{qu}}} dq_z
\int_{-\pi}^{\pi} \frac{d q_x}{2 \pi}\times
\nonumber\\
&&\times \int_{- \pi}^{\pi} \frac{d q_y}{2 \pi} \frac{1 - \cos
({\bf q_{\perp} g})}{\left(\frac{T}{\alpha \delta_{qu}(T)
T_c^{(1)}} \right)^2 n^2 + q_z^2 + \omega_{\perp}^2(T)}.
\label{S(g,T)1}
\end{eqnarray}
We use the transformation $1/a = \int_0^{\infty} e^{-a u} du$ in
Eq.(\ref{S(g,T)1}) and carry out the integrations over $q_z, q_x$
and $q_y$. For the particular case of $N_x = N$ and $N_y = 0$, the
expression for $S_{\alpha} ({\bf g}, T)$ is reduced into the
following form
\begin{eqnarray}
&&S_{\alpha}(N, T) = \frac{\alpha}{2 \sqrt{\pi}} \int_0^{\infty}
\frac{d u}{\sqrt{u}} e^{-2 u} I_0(u) [I_0(u) - I_N(u)] \times
\nonumber\\
&&\times \mathrm{erf}\left(\delta_{qu}^{- 1}(T) \sqrt{u
/2}\right) \Phi \left(u, \frac{T}{\sqrt{2} \alpha \delta_{qu}(T)
T_c^{(1)}}\right),
\label{S(g,T)2}
\end{eqnarray}
where $\Phi (u,\tau)$ is given by
\begin{eqnarray}
\Phi (u,\tau) &=& \tau \sum_{n = - \infty}^{\infty} \exp \{- u
{\tau}^2 n^2 \} \nonumber\\
&=& \left\{ \begin{array}{ll} \tau (1 + 2
e^{-u{\tau}^2}), & u>{\tau}^{-2}\\
\sqrt{{\pi}/u} + \tau, & u < {\tau}^{- 2}
\end{array} \right\};
\label{elliptictheta}
\end{eqnarray}
here $\tau$ represents the normalized temperature (see Eq.(54))
and the sum in Eq.(\ref{elliptictheta}) is also known as the
so-called {\it EllipticTheta}-function ${\sc
\theta}_3(0,\exp(-u\tau^2))$, \cite{wolfram}. Using the asymptotic
expressions for the Bessel and the error function as well as for
$\Phi (u, \tau)$ in Eq.(\ref{S(g,T)2}), we get the following
explicit expressions for $\exp \{ - S_{\alpha} (N,T)\}$
\nopagebreak
\begin{widetext}
\begin{eqnarray}
e^{- S_{\alpha}(N, T)} = \left\{
\begin{array}{ll} \delta_{qu}^{\alpha} \exp \left\{- C \frac{T}{\delta_{qu} T_c^{(1)}} + \frac{\alpha}{4 \pi}
\frac{1}{N}\right\}, & 0 \le 2 \delta_{qu}^2 < 1 < N <
\left(\frac{\sqrt{2} \alpha \delta_{qu}
T_c^{(1)}}{T}\right)^2\\
\delta_{qu}^{\alpha} \exp \left\{- C \frac{T}{\delta_{qu}
T_c^{(1)}} + \frac{C_3}{N^{1/2}} \frac{T}{\delta_{qu}
T_c^{(1)}}\right\}, & 0 \le
2\delta_{qu}^2 < 1 < \left(\frac{\sqrt{2} \alpha \delta_{qu} T_c^{(1)}}{T}\right)^2 < N\\
\left(\frac{T}{\alpha T_c^{(1)}} \right)^{\alpha} \exp \left\{-C
\frac{T}{\delta_{qu} T_c^{(1)}} +
\frac{C_4}{N^{1/2}}\frac{T}{\delta_{qu} T_c^{(1)}}\right\}, &
0 \le 2\delta_{qu}^2 < \left(\frac{\sqrt{2} \alpha \delta_{qu} T_c^{(1)}}{T}\right)^2 < 1 < N\\
\exp \left\{-C\frac{T}{\delta_{qu} T_c^{(1)}} +\frac{C_4}{N^{1/2}}
\frac{T}{\delta_{qu} T_c^{(1)}}\right\}, &
0 \le \left(\frac{\sqrt{2} \alpha \delta_{qu} T_c^{(1)}}{T}\right)^2 < 2\delta_{qu}^2 < 1 < N\\
\end{array} \right.
\label{S(g,T)F}
\end{eqnarray}
\end{widetext}
where $C_3, C_4$, and $C_5$ are again constants of order unity.


\begin{thebibliography}{99}
\bibitem{Pieirls} The ground state of a non-interacting electron
gas in an isolated $1D$ wire or in a quasi-1D system with a nested
Fermi surface is a Peierls dielectric. Suggestion of $n$-
subband in the cross-section of a pure single wire surves to
prevent the Peierls transition in the system. Disorder in the
quasi-1D system destroys also the perfect nesting, suppressing at
the same time the Peierls transition. If isolated wires contain
point-like impurities as well, the ground state becomes a dielectric
phase due to Anderson localization, and the transition becomes
superconducting-insulator phase transition. In our case, anion
disorder in the organic superconductors may not modify the Fermi
surface.
\bibitem{phonon} According to Eq.(\ref{coeff}) $|A_{\bf q}|^2
\propto (1/N)$, this approximation is valid for $N \gg 1$.
Strictly speaking, this harmonic approximation is valid for small
temperatures when $\sum_{\bf q}|A_{\bf q}|^2 \langle {\hat N}_{\bf
q} \rangle < 1 $, which means that few phonons are excited in the
system.
\end{thebibliography}
\end{document}